# Slicing Through Multicolor Space: Galaxy Redshifts From Broadband Photometry

A. J. Connolly[2], I. Csabai[1] and A. S. Szalay[1,2]

Department of Physics and Astronomy, The Johns Hopkins University, Baltimore, MD 21218

Electronic mail: ajc@skysrv.pha.jhu.edu, csabai@skysrv.pha.jhu.edu, szalay@skysrv.pha.jhu.edu

D.C. Koo[2]

University of California Observatories, Lick Observatory, Board of Studies in Astronomy and Astrophysics, University of California Santa Cruz, CA 95064

Electronic mail: koo@lick.ucsc.edu

R.G. Kron[2]

Fermi National Accelerator Laboratory, MS 127, Box 500, Batavia, Illinois 60510

Electronic mail: kron@fnal.fnal.gov

J.A. Munn[2]

Yerkes Observatory, University of Chicago, P.O. Box 258, Williams Bay, WI 53191

Electronic mail: jam@yerkes.uchicago.edu

[1]Department of Physics, Eötvös University, Budapest, Hungary, H-1088

[2]Visiting Astronomer, Kitt Peak National Observatory, National Optical Astronomy Observatories, which is operated by the Association of Universities for Research in Astronomy, Inc., under contract with the National Science Foundation.

# ABSTRACT

As a means of better understanding the evolution of optically selected galaxies we consider the distribution of galaxies within the multicolor space $U$, $B_J$, $R_F$ and $I_N$. We find that they form an almost planar distribution out to $B_J = 22.5$ and $z < 0.3$. The position of a galaxy within this plane is dependent on its redshift, luminosity and spectral type. While in the original $U$, $B_J$, $R_F$ and $I_N$ space these properties are highly correlated we can define an optimal rotation of the photometric axes that makes much of this information orthogonal. Fitting the observed spectroscopic redshifts with a quadratic function of the four magnitudes we show that redshifts for galaxies can be estimated to an accuracy better than $\Delta z = 0.05$. This dispersion is due to the photometric uncertainties within the photographic data. Assuming no galaxy evolution we derive a set of simulated galaxy fluxes in the U, J, F and N passbands. Using these data we investigate how the redshift is encoded within the broadband magnitudes and the intrinsic dispersion of the photometric-redshift relation. We find that the signal that defines a galaxy's photometric redshift is not related to specific absorption or emission lines but comes from the break in the overall shape of the galaxy continuum at around 4000 Å. Using high signal-to-noise photometric data we estimate that it is possible to achieve an intrinsic dispersion of less than $\Delta z = 0.02$.

*Subject headings:* galaxies, photometry, redshifts

## 1. INTRODUCTION

Studying the evolution of galaxies at faint magnitudes, e.g. changes in clustering and luminosity, requires detailed information on the distribution of their colors and distances. While spectra are now being obtained for galaxies to $B \leq 24$ (Glazebrook et al. 1995), the amount of time required to undertake a wide angle survey would be prohibitive. In comparison, to survey a comparable region to the same equivalent depth, in four broad band photometric bandpasses, would require two orders of magnitude less time. The uncertainties in studies of the evolution of galaxies are dominated by shot noise (statistically small samples of galaxies) rather than errors in redshift. If we could derive an estimate of the redshift of a galaxy from its photometric magnitudes then large, complete surveys could be realised.

Techniques for deriving redshifts from broadband photometry were pioneered by Baum (1962) who used nine passbands to detect the 4000 Å break in galaxies from three distant clusters. Subsequent implementations of these basic techniques have been made by Koo (1985), using four broadband photographic filters, and Loh and Spillar (1986), using nine medium bandpass filters. All applications have, however, been limited by the lack of redshift surveys to the limit of the photometric data. Consequently, redshifts had to be derived by matching the observed galaxy colors with those predicted from spectral energy distributions (SEDs) and assumed galaxy evolution models.

Given this, the results of these earlier analyses have been encouraging for intermediate redshift galaxies ($z < 0.3$), with errors of $\Delta z = 0.07$ (Koo 1985). For more distant samples the uncertainties in the derived redshifts increase to $\Delta z > 0.1$ and the error distributions are non-Gaussian. Due to these large uncertainties application of photometric redshifts to statistical studies of galaxy evolution has been limited (Caditz and Petrosian 1989).

In this paper we derive an empirical method for calculating statistical redshifts from four color photometry. For the first time we utilize both the color and magnitude information present in the galaxy catalogs, essentially treating the photometry as a very low resolution spectrum. In §2 we describe the photometric and spectroscopic data used. We show the distribution of galaxies within the multi-dimensional color space in §3 and in §4 derive a new photometric coordinate system that correlates with the physical properties of galaxies. The techniques used to estimate redshift and its accuracy are discussed in §5 and simulations using SED's to determine the intrinsic dispersion of the relation are presented in §6. We close with a discussion of the applicability of these estimated redshifts to current and future photometric surveys.

## 2. DATA

The sample for our analysis is extracted from a subset of the data in the catalogs of Koo and Kron that cover two high-galactic latitude fields, Selected Areas 57 and 68 (Munn et al. 1995). Details of the spectroscopic data can be found in the Munn et al. reference while the photometry for SA57 and SA68 is described by Kron (1980) and Koo (1986), except for later refinements (e.g. saturation corrections) as described by Munn et al.. The following is a brief description of these data.

The photometry is based on an analysis of digitized scans of sky-limited 4-m photographic plates and includes four bandpasses, designated as $U$, $B_J$, $R_F$, and $I_N$, each corresponding roughly to more standard filters: $U$, $B$, $R$, $I$, respectively. The 50% completeness level of the survey is $B_J \approx 24$ with random photometry errors smaller than 0.3 mag by $B_J \approx 23$. Colors are defined within the same aperture.

Redshifts were acquired for over 370 galaxies in the photometric sample of SA57 and SA68 galaxies from CCD spectra taken with the Mayall 4-m telescope at Kitt Peak

National Observatory. The instruments ranged from standard long-slit to multi-slit masks or multi-fiber spectrographs, and generally covered the spectral range 4500Å to 7500Å at typical resolutions of 7-15Å. Depending on the signal-to-noise ratio of the spectra and the reliability of the detected spectral features and continuum, each redshift is assigned a quality factor. The vast bulk of the redshifts were in the moderate range from z ~ 0.1 to 0.4.

From these data, we define two samples for further analysis of the relationship of multicolor photometry to redshift. The first is a redshift sample containing all 254 galaxies brighter than $B_J = 22.5$ with spectroscopic redshifts that are of high reliability, i.e. less than 10% chance of a significant error. The second is a photometric sample of 2025 galaxies that have $B_J \leq 22.5$ and that have photometric random errors less than 0.5 mag in each of the four bandpasses. Both samples have been culled of galaxies which had unreliable photometry resulting from bad scan lines; bright galaxies have had saturation corrections applied. Table 1 presents the mean photometric errors of the spectroscopic and photometric galaxy samples, as a function of bandpass. The mean redshift of the spectroscopic sample, at the limiting magnitude shown in Column 2, is given in Column 4.

## 3. THE DISTRIBUTION OF GALAXIES IN MULTICOLOR SPACE

In our subsequent analyses we consider the distribution of the multicolor fluxes of galaxies within the four space of their $U$, $B_J$, $R_F$ and $I_N$ magnitudes. When describing the variation of their fluxes we are liberal with the verbal distinction between flux and color. To distinguish between these two quantities we will use the term color index when referring to galaxy colors (e.g. $U - B_J$).

To visualize the surprisingly strong correlations between the broadband photometry and redshift we plot the distribution of the spectroscopic sample of galaxies ($B_J < 22.5$) in the three space $U$, $B_J$ and $R_F$ (Figure 1). The redshift of each galaxy is encoded in the color of its point; blue corresponds to a redshift of zero and red to a redshift of 0.5. The color table has been chosen such that each band of color is equivalent to a redshift interval of approximately 0.1. Panels (a), (b) and (d) show different perspectives of the three dimensional distribution. These view points are, (a) face on to the distribution, (b) parallel to the flux direction (i.e. the (1,1,1) vector in $U$, $B_J$, and $R_F$ space) and (d) perpendicular to the $B_J$ axis (giving an edge on view of the distribution). Panel (c) provides a schematic of the distribution that illustrates the change in flux and color of a galaxy as a function of its redshift.

From the three perspectives we see that the galaxies form an almost planar distribution in $U$, $B_J$ and $R_F$ space. They occupy less than 4% of the available parameter space. We can, therefore, describe the data in terms of a subspace of the $U$, $B_J$ and $R_F$ magnitudes. The dimensions of this subspace are 6.5 magnitudes in the flux direction, 3 magnitudes orthogonal to the flux vector (i.e. the range of color) and less than 1 magnitude in thickness. There is evidence for some curvature within the planar distribution (see Figure 1b) but the data are essentially two dimensional.

One can easily see that the curvature within the distribution is inherent. The bands $UB_JR_FI_N$ form a roughly logarithmic sequence of central wavelengths. If the galaxy distribution at low redshifts ($z < 0.4$) forms a roughly planar distribution in $UB_JR_F$, then at higher redshifts ($z > 0.4$), the galaxy SED is shifted about one band up. What used to be in the $U$ band is shifted to $B_J$, what was in $B_J$ is now in $R_F$, etc. For no evolution, the geometric distribution of galaxies should remain the same, therefore the galaxies have to form a planar distribution, but now in $B_JR_FI_N$ rather than $UB_JR_F$. This is equivalent to a 90 degree rotation of the axes, which of course takes place continuously. Thus, curvature must be present in the multicolor distribution.

The position of a galaxy within the planar distribution is dependent on its redshift, luminosity and spectral type. If we remove the redshift effect by only considering those galaxies at a given distance (i.e. those galaxies with points of the same color in Figure 1) then they would form an iso-redshift plane (or 'slab') in the $UB_JR_F$ space. We illustrate this in Figure 1c. The natural coordinates within this iso-redshift slab are the luminosity and intrinsic color. The red edge of this plane would be skewed towards brighter magnitudes because red galaxies are intrinsically more luminous than blue galaxies (Binggeli et al. 1987, Loveday et al. 1992, Bershady 1995). This effect is visible in the $UB_JR_F$ data, where the galaxies at a given redshift in Figure 1b lie in skewed slabs.

As these galaxies are redshifted, their magnitudes dim (simply the $1/r^2$ effect) and their colors change (i.e. K corrections). These two components shift the planar distributions both in the flux and color directions. The result is a 'shifted stack of slabs' (of constant redshift) passing through the multicolor distribution (see Figure 1c). As we shall show, our ability to visually distinguish, to an accuracy of $\Delta z = 0.1$, between galaxies of different redshift within the $UB_JR_F$ plots manifests as a means of estimating the redshift of a galaxy from its broadband colors.

A further consequence of this interpretation is the potential for deriving a coordinate system, directly from photometric observations, that relates to the physical properties of a galaxy (e.g. redshift, SED type and intrinsic luminosity) rather than its photometric magnitudes.

### 3.1. Dimensionality of the galaxy distribution

We can quantify the approximate geometry of the data within the four-space of $UB_JR_FI_N$ by deriving their Hausdorff or fractal dimension. This relates the number of neighboring points, $M$, within a distance $R$ (Hausdorff, 1919). If we consider a one dimensional distribution of data then the number of neighbors scales proportionately to the distance. For two dimensional data $M$ scales as $R^2$, and for the general case

$$M(<R) \propto R^D. \tag{1}$$

The exponent, $D$, is the derived dimensionality of the given distribution.

From the spectroscopic sample we calculate the Hausdorff dimension of the data using four different distance metrics, the 3D color indices $(R_3)$, the $UB_JR_FI_N$ magnitudes $(R_4)$, magnitudes and redshift $(R_5)$, and color and redshift $(R_{4'})$ as

$$R_3 = [\Delta(U-B_J)^2/2 + \Delta(U+B_J-R_F-I_N)^2/4 + \Delta(R_F-I_N)^2/2]^{\frac{1}{2}},$$
$$R_4 = [\Delta U^2 + \Delta B_J^2 + \Delta R_F^2 + \Delta I_N^2]^{\frac{1}{2}},$$
$$R_5 = [\Delta U^2 + \Delta B_J^2 + \Delta R_F^2 + \Delta I_N^2 + \Delta(5\log z)^2]^{\frac{1}{2}},$$
$$R_{4'} = [[\Delta(U-B_J)^2/2 + \Delta(U+B_J-R_F-I_N)^2/4 + \Delta(R_F-I_N)^2/2 + \Delta(5\log z)^2]^{\frac{1}{2}}.$$

In order to create a flux independent 3D subspace, but preserve Euclidian distances in magnitudes, we have rotated the $UB_JR_FI_N$ space to align one of the axes with the flux direction, along $(U+B_J+R_F+I_N)$. Two other axes were chosen to be parallel with $(U-B_J)$ and $(R_F-I_N)$, thus the fourth axis is along $(U+B_J-R_F-I_N)$. The redshift simply causes a dimming of $5\log z$ magnitudes; this is used when we include the redshift in our dimensionality analysis.

The number of neighbors as a function of these distances is shown in Figure 2. Each of the Hausdorff lines in Figure 2 deviates from a straight line for large $R$. This is due to the boundaries of the data – as $R$ increases beyond the largest extent of the data (6.5 magnitudes), all points are inside and the dimensionality tends towards zero. All other characteristic sizes (like the width and the thickness of the 'wedge', about 3 and 1 magnitudes, respectively) show up as smooth breakpoints in the $M(<R)$ curve. Because of this the dimensionality we derive from a linear fit to the Hausdorff curves is sensitive to the range of magnitudes or color we fit over. As we are interested in the intrinsic dimensionality of our data, and would like to stay away from the regime where edge effects become important, we fit a linear relation in $\log M$ vs $\log R$ to all samples containing flux or redshift between $0.1 < R < 1$ magnitudes, and we fit the color relation between $0.025 < R < 0.4$ magnitudes (see Table 2). On these scales the uncertainty in the derived dimensionality is approximately 0.1.

If we consider only the color indices of the galaxies, the dimension is $D_3 = 1.9$. Including the additional dimension of redshift to the colors, the dimensionality of the system remains essentially constant, $D_{4'} = 1.8$. Redshift is, therefore, strongly correlated with the relative colors, at least locally. This does not, however, mean that we can easily decode the redshift of a galaxy from just its colors as the variation of color with redshift may be degenerate, i.e. galaxies of different spectral type may have similar colors at different redshifts (Koo 1985). If we add flux, rather than redshift to the colors, there is an increase to $D_4 = 2.3$. Adding both total flux and redshift we obtain $D_5 = 2.5$.

All these derived dimensions are a natural consequence of the intrinsic properties of galaxies. The fact that the galaxies are located in a subspace with fewer dimensions than the number of photometric bands implies that there is redundant information present. This is consistent with results found by Connolly et al. (1995) who show that galaxy spectral energy distributions can be reconstructed from a one-parameter linear combination of three orthogonal spectra. This sequence provides the first intrinsic dimension for the multicolor data. Redshift, via the K-correction adds another nontrivial dimension, and thus we can understand why the dimensions $D_3$ and $D_{4'}$, with colors and the tightly correlated redshift are so close to 2. Luminosity is an additional intrinsic parameter, which is known to correlate with the SED type. The measured fluxes are further correlated with redshift. This partially correlated nature of the fluxes with both colors and the redshift accounts for the fractional increase in the dimensionality, when flux is mixed with colors and/or redshift.

## 3.2. Redundancy of dimensions

If the photometric properties of a galaxy are described by a subspace with fewer dimensions than the original four color photometric system then the $UB_JR_FI_N$ magnitudes carry redundant information. We illustrate this by constructing one of the magnitudes from the other three. Using the spectroscopic sample we derive a simple linear fit of $UB_JR_F$ to the $I_N$ magnitude. We find,

$$I_N = 0.233 + 1.338R_F - 0.377B_J + 0.007U. \tag{2}$$

The standard deviation about this fit is 0.089 magnitudes.

The $I_N$ magnitude is most strongly correlated with the $R_F$ magnitude. This is expected as the spectral energy distributions of normal (i.e. non starburst) galaxies are a continuous monotonic function. If we exclude the contribution of the $U$ band to the fit the estimated $I_N$ magnitude is parameterized by a zero point (determined from the $R_F$ magnitude) and a color term ($B_J$-$R_F$). The slope of the galaxy spectral energy distribution in the near-infrared is, therefore, correlated with the spectral break in the continuum around 4000 Å.

We apply the relation derived from the 254 galaxies with spectroscopic redshifts to the full photometric sample. Figure 3 shows the estimated $I_N$ band magnitudes for these data compared with their measured $I_N$ magnitudes. The line drawn in Figure 3 is not a fit to the data, it is simply a line with a slope of unity and passing through the origin. At $I_N$=20 the dispersion about this line is 0.393 magnitudes. This compares with an estimate of the photometric uncertainties within the data of between 0.2 (SA57) and 0.5 (SA68) magnitudes (Koo 1986). The correlation between the estimated and measured $I_N$ magnitudes is, therefore, as tight as the photometric uncertainties. This is not as surprising as it first seems. The $B_J$ and $R_F$ magnitudes have smaller photometric errors than those for the $I_N$ and $U$ bandpasses. If, as we have shown, the data are of intrinsically lower dimensionality than the full $UB_JR_FI_N$ space we can use the highest quality data to reconstruct those with larger photometric errors.

## 4. THE NATURAL MULTICOLOR COORDINATE SYSTEM

The distribution of the galaxies in the 4-dimensional $UB_JR_FI_N$ space is quite planar. The finite thickness of this plane carries significant physical information, not just random scatter. We have seen above that the intrinsic dimensionality of this space is lower than the four color space, somewhere between two and three. Here we attempt to find a rotation of this 4-dimensional space, such that the resulting coordinates have a closer relation to the intrinsic properties defining a galaxy (redshift, SED type and luminosity). One way to arrive at such orthogonal axes is by using the principal components of the distribution. This will be one of the systems that we will discuss. The other system will be derived by directly correlating the galaxy distribution with redshift and flux.

There are several natural directions in the 4-dimensional space. One of them is the normal to the plane of the galaxy distribution. We have seen that the $I_N$ magnitude was quite well fitted by a linear function of $UB_JR_F$. This fit defines the 4D normal vector to the plane of the galaxies, which we denote by $\hat{n}$. A linear fit to the redshift of the galaxies in 4D space yields the normal vector $\hat{z}$, along the direction of maximal correlation. These two vectors are not orthogonal to each other, but they define a 2D subspace. We define the normal vector $\hat{s}$, to be within this subspace, normal to $\hat{z}$, i.e. uncorrelated with redshift. Within the 2D subspace, orthogonal to these two vectors, we choose another vector $(\hat{f})$, to be maximally correlated with the total flux, along the $(1, 1, 1, 1)$ direction. The remaining direction ($\hat{t}$) is now fully determined. We have created a coordinate system, in which (at least in linear order) the redshift is correlated with only one of the axes ($\hat{z}$), flux is correlated with another ($\hat{f}$). The direction $\hat{t}$ is within the plane of the galaxy distribution, orthogonal to redshift and flux, thus it is correlated with the SED type. The fourth direction ($\hat{s}$), is orthogonal to all the above, and represents the scatter away from the plane of the galaxies, uncorrelated with all the above. The 'edge-on', 'face-on' and 'down' views of Figures 1a-d closely correspond to these coordinates, when taken in the 3D subset of $UB_JR_F$ colors.

$$\hat{n} = (0.05, -0.30, 0.78, -0.55)$$

$$\hat{s} = (-0.09, 0.07, 0.72, -0.68)$$
$$\hat{z} = (0.32, -0.88, 0.30, 0.18)$$
$$\hat{t} = (0.79, 0.06, -0.37, -0.48)$$
$$\hat{f} = (0.52, 0.46, 0.50, 0.52)$$

These directions should be compared with the principal coordinates of the $UB_JR_FI_N$ color space. We construct the principal coordinates within the center of mass of the distribution (i.e. we subtract the mean). The four components, ordered by eigenvalue are given below.

$$\hat{p}_1 = (0.76, 0.47, 0.32, 0.33)$$
$$\hat{p}_2 = (-0.47, 0.69, -0.33, 0.43)$$
$$\hat{p}_3 = (-0.33, -0.32, 0.62, 0.63)$$
$$\hat{p}_4 = (0.30, -0.46, -0.63, 0.56)$$

The first principal component contains 97.98% of the power (or information) within the multicolor system (defined as the ratio of the eigenvalue of a particular principal component to the sum of all eigenvalues). By the second component the power has fallen to 1.98% and 0.04% by the third. Describing the data in terms of a planar distribution is a natural consequence of the principal component analysis.

The correlation between the principal coordinates and the coordinate system derived using a priori information is found by projecting the principal vectors onto the latter system. We reorder the principal coordinates to form the matrix,

|           | $\hat{s}$ | $\hat{z}$ | $\hat{t}$ | $\hat{f}$ |
|-----------|-----------|-----------|-----------|-----------|
| $\hat{p}_4$ | 0.90    | −0.40     | −0.17     | 0.08      |
| $\hat{p}_2$ | 0.44    | 0.78      | 0.42      | −0.13     |
| $\hat{p}_3$ | −0.03   | −0.47     | 0.82      | −0.32     |
| $\hat{p}_1$ | −0.02   | −0.02     | 0.35      | 0.94      |

The principal coordinates are very similar to those derived from the physical properties of galaxies. The first principal component corresponds to the flux direction (lying less than 20° away). It also contains a small component in the $\hat{t}$ direction (intrinsic luminosity and spectral type are strongly correlated). The second component correlates strongest with the redshift direction and the third with the spectral type. It is important to note that we include no redshift information when constructing the principal coordinates and yet the separation of redshift, type and flux are a natural consequence.

A further comparison can be made with the $UB_JR_FI_N$ directions derived by Koo (1986), who used only sums and differences of the broadband magnitudes. The vectors presented below are not normalized.

$$\hat{Z} = (-1, 2, -1, 0)$$
$$\hat{T} = (2, 0, -2, 0)$$
$$\hat{F} = (1, 1, 1, 1)$$

The flux and redshift directions are almost identical to those found when rotating the photometric axis using our a priori information. The spectral type direction $(\hat{T})$ initially appears to be significantly different until we recall the almost one-to-one correlation between $R_F$ and $I_N$ magnitudes. If we substitute the linear correlation of Eqn 2 the directions are again identical.

From the analyses above we can define a grid in multicolor space that separates the effect of redshift, luminosity and spectral type. This is the first step in analyzing galaxy distributions in a natural coordinate system rather than that imposed by an arbitrary photometric system. The first order alignment (i.e. linear) we show here is not the best representation of true distribution of galaxy magnitudes. As we have seen, the multicolor plane which galaxies occupy has an intrinsic curvature (due to the redshifting of the galaxy spectra through the different bandpasses). By applying small distortions to this linear grid we can improve the fit to the data. This will be addressed in a subsequent paper.

## 5. PHOTOMETRIC ESTIMATES OF REDSHIFT

Previous attempts to assign redshifts to galaxies based on their broadband photometry have made use of the variation in the color of a galaxy as a function of its redshift (Koo 1985). By using model spectral energy distributions (e.g. Bruzual and Charlot 1993) and assuming a particular evolutionary scenario the changes in color can be traced for different spectral types. While this approach has proved successful for estimating the redshifts of local galaxies ($z < 0.25$) problems occur at higher redshifts. The change in color as a function of redshift becomes degenerate; high redshift intrinsically blue galaxies can have similar colors to low redshift intrinsically red galaxies.

By incorporating the magnitude of a galaxy we can remove much of this degeneracy. Using an empirical approach we determine the change in the multicolor fluxes of galaxies as a function of their redshift. We use those data with accurately measured spectroscopic redshifts and $B_J < 22.5$ to define the photometric redshift relation. This sample has a median redshift of 0.25 and extends out to a redshift of 0.6.

## 5.1. Linear fits

A linear correlation between the $UB_JR_FI_N$ magnitudes and redshift was derived for the spectroscopic sample of galaxies. We do not include the formal photometric errors for the individual galaxies when determining our fits, nor do we correct for the effect of selection biases that may be present within the data (e.g. Malmquist bias). The linear regression is given by,

$$Z = -0.941 - 0.147U + 0.412B_J - 0.138R_F - 0.084I_N \qquad (3)$$

The rms dispersion about this fit is $\sigma_z = 0.057$. We introduce the reduced $\chi^2$ of this fit in order to compare how well the different order fits represent the distribution of the data. We derive a normalization for the $\chi^2$ by assuming an error of 0.1 magnitudes in each passband (consistent with the errors quoted in Table 1). This normalization is somewhat arbitrary as we do not include the photometric errors in our fits. We, therefore, only consider the relative change in $\chi^2$. For a linear fit $\chi^2$ is 3.31

From a simple linear correlation between multicolor fluxes and redshift we can derive redshifts accurate to approximately 20%. The residuals about this fit correlate with redshift. From Figure 1 we see that galaxies occupy a curved plane in the multicolor space. By applying a linear fit we are deriving the tangent to this plane (we intercept the plane at the center of mass of the distribution; z=0.25). Galaxies that lie away from the median redshift have a systematic offset from the estimated redshift. For z=0.5 this is approximately $\Delta z = 0.05$.

## 5.2. Quadratic fits

We repeat the analysis above using a quadratic form for the fitting function. Again we weight all data points equally, irrespective of their photometric errors. The quadratic fit to the data is given by,

$$
\begin{aligned}
z \;=\; & 0.396 + 0.121U - 0.0990B_J - 0.868R_F + 0.803I_N \\
& -0.346UB_J - 0.452UR_F + 0.0914UI_N \\
& +1.256B_JR_F - 0.263B_JI_N + 0.169R_FI_N \\
& -0.008246U^2 - 0.636B_J^2 - 0.485R_F^2 - 0.0177I_N^2
\end{aligned}
$$

The formal dispersion about this fit is $\sigma_z = 0.047$, a decrease of 20% from the linear fit. The $\chi^2$ for a quadratic fit reduces to 2.32.

The correlation between estimated and measured redshift is shown in Figure 4. The line drawn in Figure 4 passes through the origin and has a slope of unity (it is not a fit to the data). There is a strong correlation between the estimated and measured redshifts through the full redshift range of the data. Beyond z=0.45 the photometric redshifts are systematically lower than the measured redshifts (the effect is small, at the 10% level). This is due to the effect of Malmquist bias within the galaxy sample. At the magnitude limit of the sample photometric errors cause intrinsically more distant galaxies to be scattered into the sample. The photometric-redshift relation, consequently, underestimates their distances. The effect for these data is small but the presence of such a bias reiterates the need for accurate photometric data (the number of galaxies scatter into the sample correlates with photometric error).

The error in the estimated redshift is shown for each galaxy as a function of its $B_J$ magnitude in Figure 5a. The uncertainty increases as a function of magnitude (or photometric error). Up to a magnitude limit of $B_J = 21$ there is no systematic deviation from a zero mean error. Beyond $B_J = 21$ the effect of Malmquist bias becomes apparent.

The error in the photometric-redshifts decreases as a function of $U - B_J$ color (see Figure 5b). For galaxies with $U - B_J < 0$ the dispersion in the relation is nearly 40% greater than for those galaxies with redder colors. Blue galaxies have intrinsically flatter spectra than red galaxies. If the source of the signal for the photometric redshift comes from the continuum shape then it is expected that blue galaxies will have greater uncertainties in their estimated redshifts. The photometric uncertainty of the data, however, also increases for blue galaxies. At the $B_J$ magnitude limit of the data ($B_J$=22.5) the $U$ band photometric error is approximately 0.4 magnitudes. This may contribute, in part, to the error in the estimated redshifts. To discriminate between these effects we require simulations.

### 5.3. Higher Order fits

To determine the complexity of the photometric-redshift relation we fit third and fourth order polynomials to the data. For the third order fit the dispersion decreases to 0.042 and the reduced $\chi^2$ decreases to 2.07. This represents an increase in accuracy of approximately 10%. For the fourth order fit the dispersion decreases to 0.038 but the change in the $\chi^2$ remains essentially unchanged at 2.01. That the $\chi^2$ and dispersion decrease as we go to higher order fits indicates that there is curvature within the relation. As we shall see in our simulations there are regions of redshift space that require higher order terms to fully map the change in flux as a function of redshift. The fact that the $\chi^2$ is reduced only slightly when we move from a third to a fourth order fit suggests that the curvature in the relation is not in all of the dimensions (i.e. the planes that define luminosity and spectral type may change slowly as we move to higher redshift).

## 6. SIMULATIONS

As seen above the errors within the photometric-redshift relation can, in part, be explained by the photometric uncertainties present within the data sets (particularly the uncertainties in the $U$ band data). In order to estimate the intrinsic dispersion within the relation and to determine which bandpasses provide the most information we derive simulated galaxy catalogs using the spectral energy distributions of Rocca-Volmerange and Guiderdoni (1988).

Assuming a no evolutionary model, i.e. excluding the effects of star formation or emission lines, and using the 15 Gyr spectral energy distributions we compute two simulated catalogs, one with a signal-to-noise ratio of the original $U$, $B_J$, $R_F$ and $I_N$ photographic data and a second with a signal-to-noise ratio of 10, at a limiting magnitude of $U$=24, $B_J$=24, $R_F$=23.5, $I_N$=22.5 (equivalent to a multicolor CCD survey currently in progress). Each catalog was derived assuming the following parameters. We use the n(z) distribution given by Koo and Kron (1992) for galaxies with $B_J < 26$ and define $q_0$=0.5 and $H_o = 100$ Mpc$^{-1}$ km s$^{-1}$. The morphological distribution is that found by Giovanelli et al.(1986) in their survey of the Perseus-Pisces supercluster and the luminosity functions adopted are those of Loveday et al. (1992) with M$^* = -19.71$ for E and S0 galaxies and M$^* = -19.40$ for spirals.

## 6.1. The intrinsic photometric-redshift relation

We verify that our simulations are a fair representation of the optical data by fitting a quadratic photometric-redshift relation to the sample with photometric errors identical to those given by Koo (1986). The correlation between the estimated and spectroscopic redshifts is shown in Figure 6a. The dispersion about a one-to-one correlation is $\sigma_z = 0.045$, almost identical to the $\sigma_z = 0.047$ derived from the photographic data.

In Figure 6b we show the correlation between the estimated and true redshift for the high signal-to-noise simulation. The data have been fitted by a quadratic function in $UB_JR_FI_N$ for galaxies with $B_J < 24$. The dispersion about this fit is $\sigma_z = 0.044$. It is remarkable that between a redshift of 0.0 and 0.8 we can fit the change in flux as a function of redshift to such a high accuracy with just a quadratic relation. We can improve on this fit by reducing the redshift range over which we fit. We fit a quadratic spline to the photometric redshift relation, essentially fitting tangent curves to the curved relation. With a node separation of $\Delta z = 0.2$ the $\sigma_z$ decreases to 0.018.

From §3, we know that the distribution of galaxies in multicolor space has an intrinsic curvature. By fitting a quadratic function to the photometric-redshift relation we are trying to represent this curvature with a smoothly varying function. At $z = 0.4$ there is evidence for a kink in the photometric redshift relation (see Figure 6b). This occurs when the galaxy colors rotate by 90° in the four color space with each restframe filter redshifting into its adjacent bandpass. The rapid change in galaxy color as a function of redshift is not well matched by the slowly varying quadratic function. Adopting the quadratic spline or an iterative approach where we fit the general shape of the photometric-redshift relation across the full redshift range and then refit over smaller redshift intervals is equivalent to the perturbation of the multicolor grid proposed in §4.

At redshifts greater than unity the photometric-redshift relation breaks down for the optical bandpasses. The restframe spectral features have redshifted out of the optical bandpasses (e.g. the break in the continuum at 4000 Å present in galaxy spectra is redshifted into the $I_N$ band). We are therefore considering different spectral regions when we compare the $UB_JR_FI_N$ colors of local galaxies with those at high redshift. To extend the photometric-redshift relation we require that the bandpasses extend into the near-infrared.

## 6.2. Determining the source of the redshift signal

We determine which and how many colors are required to define the photometric-redshift relation by fitting the high signal-to-noise simulations with a subset of the $UB_JR_F$ and $I_N$ filters. Table 3 shows the $\sigma_z$ for each combination of filters used in the quadratic fits. For $0.0 < z < 0.8$ the dispersion about the fit is 0.044 when using all four colors. If we exclude either the $B_J$, $R_F$ or $I_N$ colors from the fit the dispersion increases by approximately 20%. If, however, we exclude the $U$ band the dispersion increases by 60%. Clearly it is the variation of $U$ flux that dominates the accuracy of our photometric redshifts. Equivalently it is the shape of the galaxy spectrum in the $U$ band that provides much of the signal for determining the redshift of a galaxy from its broadband colors.

As noted above, when we consider galaxies with $z > 0.4$ the features present in a particular spectrum are redshifted into the adjacent bandpass. To verify which restframe spectral features define the accuracy of our photometric-redshift relation we fit the relation between a redshift of 0.0 and 0.4. The dispersion about this fit, for all four bandpasses, is 0.016. If we exclude the $R_F$ or $I_N$ filters the dispersion is essentially unaltered. If we remove either the $U$ or $B_J$ bandpasses $\sigma_z$ more than doubles. For redshifts greater than 0.4 the $U$ band is no longer dominant in defining the accuracy of a photometric-redshift. The dispersion increases by only 14% when we exclude it from the fits. In contrast excluding either of the red bandpasses increase the dispersion by about 50%. This is consistent with the local fits – the feature present in the restframe $U$ band has been redshifted into the longer bandpasses.

To determine the number of required colors to define the redshift of a galaxy we consider only the fits for $z < 0.4$. This is to ensure that we work in the restframe bandpasses. In Table 4 we show the $\sigma_z$ as a function of the number of bandpasses used in a fit (all fits are quadratic). We add filters in order of increasing effective wavelength. For the first three filters the dispersion about the relation decreases as we increase the number of filters. Adding the $I_N$ band to the fit does not, however, significantly decrease $\sigma_z$. As we showed previously, the photometric data are of an intrinsically lower dimensionality than the number of photometric axes. The luminosity, redshift and spectral type are encoded in three, almost orthogonal, dimensions. Our ability to define the redshift of a galaxy using only three colors is, therefore, a natural consequence. Extending the fits to higher redshift ($z > 0.4$) we could again fit the photometric redshift relation with three colors ($B_J R_F I_N$). However, fitting the relation from $0.0 < z < 0.8$ requires additional information (i.e. bandpasses) to account for the curvature as we rotate from one photometric system into the next.

### 6.3. Defining redshift as a function of spectral type

If we consider the photometric-redshift relation as a function of spectral or spectral type we see that there exists a correlation between the dispersion of the relation and the spectral type of a galaxy. Table 5 shows the dispersion in the photometric-redshift relation fitted to the high signal-to-noise simulation ($B_J < 24$) as a function of spectral energy distribution (we give both the morphological type of the spectral energy distribution and its $B - V$ colors). For intrinsically red galaxies the dispersion is a factor of two smaller than the over all dispersion. This is consistent with the findings of (Butchins 1981) who showed that redshifts of elliptical galaxies could be determined to an accuracy of 0.03 from two color photometry. As we move to progressively bluer galaxy types the dispersion increases. The dispersion of the relation is dominated by that of the bluest galaxies. While all spectral types seem to follow the same general photometric redshift relation fitting the separate spectral types individually may improve on the accuracy of our fits.

## 7. DISCUSSION

As we have shown, it is possible to derive the redshift of a galaxy to an accuracy of $\Delta z = 0.05$ using only four broadband colors. For medium resolution spectra we derive redshifts by identifying spectral features as they are redshifted to longer wavelength bins. If we consider the broadband photometry in terms of a low resolution spectrograph (with a resolving power of $\sim 5$) then the principal question is what is the spectral feature (or signal) that we are using to determine the redshift of a galaxy? With a resolution of approximately 1000 Å for each band pass we are clearly not identifying the passage of particular emission or absorption features through a bandpass.

The signature we detect is the effect of passing the overall shape of the continuum through the different bandpasses as the redshift increases. The break in the general continuum shape at 4000 Å for normal galaxies acts in the same way as a spectral feature is used to define a redshift for higher resolution data (it acts as a discontinuity within the spectrum). That the amplitude of this break decreases as we go to progressively bluer galaxies (e.g. irregulars) explains why the dispersion within the estimated redshifts increases for late type galaxies (see §5). In blue galaxies (i.e. those undergoing strong star formation) the rise in the UV emission shortward of 3500 Å provides a discontinuity analogous to the break at 4000 Å (Connolly et al. 1995).

In their restframe the distribution of galaxies can be described by three filters (see §3.1 and §6.2). Additional bandpasses do not add to the accuracy to which we can estimate redshifts, for galaxies with $z < 0.4$. As we move to higher redshifts curvature is added to this three dimensional plane (i.e. the restframe colors rotate into the longer wavelength filters). We, therefore, require an additional component (or color) to characterize this curvature. For intermediate redshift galaxies ($z < 0.4$) the $U$ and $B_J$ bands define the accuracy to which a redshift can be determined. As we move to higher redshifts the continuum break around 4000 Å moves into the $B_J$ and $R_F$ bands and these become the significant bandpasses (see Table 3). As a rule of thumb, in order to utilize broadband photometry for an estimate of galaxy redshifts we require filters that straddle the 4000 Å spectral feature and a filter longward of the break that acts as a fiducial mark. With optical bands, the simulations indicate a breakdown in the simple quadratic form of the photometric-redshift relation (i.e. the feature has moved out of our filter system) by redshifts of unity.

While our estimates of the intrinsic dispersion of the photometric-redshift relation must be considered lower bounds (we do not attempt to simulate the effect of internal extinction or galaxy evolution in our models), an accuracy of even $\sigma_z = 0.05$, as determined from the photographic data, opens numerous opportunities for study of the distribution and evolution of galaxies. Applying the photometric-redshift relation to the photometric sample of galaxies we can determine the luminosity function with unprecedented accuracy - studying both the local distribution and its evolution (Subbarao et al. 1995). Other applications include the detection of clusters of galaxies (Koo et al. 1988) and the evolution of the angular correlation function as a function of redshift (rather than magnitude).

We can extend our analysis from simply defining an estimated redshift to a galaxy by considering the data within its natural coordinate system. §4 showed that we can define orthogonal axes within the four color space that maximize the separation redshift, flux and spectral information. By considering the distribution of galaxies within this multicolor space we can trace the spectral evolution of the data as a function of magnitude and redshift. An obvious goal of these techniques is their application to the Sloan Digital Sky Survey. Extending the number of galaxies with distance, luminosity and spectral type information from the spectroscopic redshift survey limit of $B \sim 18.3$ to the photometric limit of $B \sim 23.5$ would increase the sample of galaxies available for study from $10^6$ to $510^7$.

## 8.   CONCLUSIONS

Using an empirical approach, we define the distribution of galaxies within the multicolor space of $UB_J R_F I_N$. We find that,

(1) Galaxies occupy a 2 or 3 dimensional subspace of the original four color space. The position of a galaxy within this subspace is determined by its redshift, luminosity and spectral type. We derive an optimal rotation of the original photometric axes that produces axes that correlate (in a linear sense) with these physical properties of the galaxies.

(2) Fitting a quadratic function to the $UB_J R_F I_N$ photographic magnitudes we can estimate galaxy redshifts to an accuracy of $\Delta z < 0.05$ (for $B_J < 22.5$).

(3) The break in the overall galaxy continuum (at around 4000 Å) provides the signal from which photometric redshifts can be determined in the optical bandpasses.

(4) The dispersion about this relation appears to be dominated by the photometric uncertainties in the data. From high signal-to-noise simulations which assume no evolution in the galaxy populations, i.e. excluding the effects of star formation, we show that it may be possible to estimate galaxy redshifts to an accuracy of 0.02 out to a redshift of $z < 0.8$.

We thank Robert Brunner, Michael Vogeley and Gyula Szokoly for helpful discussions

and Matt Bershady, Steve Majewski, and John Smetanka for help with the photometry and spectroscopy of the SA57 and SA68 regions. AJC and AS acknowledge partial support from NSF grant AST-9020380, an NSF-Hungary Exchange Grant, the US-Hungarian Fund, the Seaver Foundation and the Sloan Digital Sky Survey. DCK acknowledges support from the NSF grant AST 88-58203.

---

Table 1.  Mean photometric errors within the U, $B_J$, $R_F$, $I_N$ sample of galaxies

| Bandpass | Limiting Magnitude | $\Delta m$ | $\bar{z}$ |
|:---:|:---:|:---:|:---:|
| $U$ | 23 | 0.48 | 0.39 |
| $B_J$ | 22 | 0.08 | 0.37 |
| $R_F$ | 21 | 0.09 | 0.44 |
| $I_N$ | 20 | 0.26 | 0.43 |

Table 2.  Hausdorff dimensionality of galaxies in $U\ B_J\ R_F\ I_N$ space

| | Flux | Flux and Redshift | Color | Color and Redshift |
|:---:|:---:|:---:|:---:|:---:|
| $D$ | 2.3 | 2.5 | 1.9 | 1.8 |

Table 3.  Dispersion of photometric-redshifts as a function of bandpass

| | $B_JR_FI_N$ | $UR_FI_N$ | $UB_JI_N$ | $UB_JR_F$ | $UB_JR_FI_N$ |
|:---:|:---:|:---:|:---:|:---:|:---:|
| $0.0 < z < 0.8$ | 0.070 | 0.050 | 0.059 | 0.057 | 0.044 |
| $0.0 < z < 0.4$ | 0.038 | 0.037 | 0.017 | 0.018 | 0.016 |
| $0.4 < z < 0.8$ | 0.050 | 0.052 | 0.070 | 0.061 | 0.043 |

Table 4.   Dispersion of photometric-redshifts as a function of number of bandpasses

| | $U$ | $UB_J$ | $UB_JR_F$ | $UB_JR_FI_N$ |
|---|---|---|---|---|
| $\sigma_z$ | 0.055 | 0.041 | 0.018 | 0.016 |

Table 5.   Dispersion of photometric-redshifts as a function of spectral type

| | E | S0 | Sa | Sb | Sc | Sd | Im |
|---|---|---|---|---|---|---|---|
| $B - V$ | 0.91 | 0.81 | 0.75 | 0.67 | 0.56 | 0.46 | 0.33 |
| $\sigma_z$ | 0.022 | 0.025 | 0.033 | 0.036 | 0.039 | 0.040 | 0.048 |

Fig. 1.— The distribution of galaxies within the three color space $U$, $B_J$ and $R_F$ is shown for the sample of galaxies derived from the spectroscopic redshift surveys of Koo and Kron. The redshift of each galaxy is encoded by the color of its data point, blue corresponds to z=0 and red to z=0.5. The color table is set so that each color maps onto an interval of 0.1 in redshift. Panels (a), (b) and (d) show three orthogonal perspectives of the data. Panel (c) shows a schematic of the distribution. The position of a galaxy within the three color space is determined by its redshift, luminosity and spectral type. For a given redshift the data form thick slabs in the $UB_JR_F$ space. Redshifting the galaxies moves these slabs through the color space (due to dimming and K corrections).

Fig. 2.— The Hausdorff or fractal dimension of the $U$ $B_J$ $R_F$ and $I_N$ spectroscopic data is shown for the flux, color, flux and redshift and color and redshift distance metrics. The intrinsic dimensionality of the data is given by the gradient of the correlation between the log of the number of neighbors within a distance D. For each of the distance metrics the data are found to exist in a subspace of the original $UB_JR_FI_N$ multicolor space (with dimensionality < 3).

Fig. 3.— In Figure 3a we fit the $I_N$ magnitude of the spectroscopic sample of galaxies to a linear combination of $UB_JR_F$ magnitudes. The line showing the linear correlation with zero intercept is not a fit to the data. The dispersion about this line is 0.089. Figure 3b shows the correlation between estimated $I_N$ magnitude and measured $I_N$ magnitude for the photometric sample (we use the fit derived from the spectroscopic sample to derive the estimated magnitudes). At $I_N$=20 the dispersion is 0.393 magnitudes.

Fig. 4.— Fitting a quadratic relation to the four magnitudes $UB_J R_F I_N$ we derive a photometric-redshift for the spectroscopic sample of galaxies. The correlation between the measured ($Z_s$) and estimated ($Z_e$) redshifts is shown. The dispersion about this relation is 0.047 for galaxies to $B_J$=22.5. The under estimate of galaxy redshifts at faint magnitudes is due to Malmquist bias (the photometric uncertainties scatter more distant galaxies into the sample).

Fig. 5.— Figure 5a shows the correlation of error in redshift ($Z_s - Z_e$) as a function of the $B_J$ magnitude. The dispersion in the redshift uncertainty correlates with $B_J$ magnitude (and, therefore, photometric error). There is, however, no systematic deviation from linearity as a function of magnitude. Figure 5b shows the redshift errors as a function of $B_J$-$R_F$ color. Blue galaxies have a systematically larger photometric redshift error.

Fig. 6.— The correlation between the estimated redshift and true redshift is shown in Fig 6a for the simulated galaxy sample with errors identical to those of Koo (1986). The dispersion of 0.045 is almost identical to the value of 0.047 derived from the photographic data. Fig 6b shows the correlation between estimated and true redshift for a sample of galaxies with signal-to-noise of 10 at $B_J = 24$. The dispersion about this fit is 0.044 out to a redshift of $z < 0.8$. By fitting the data over a range of $\Delta z = 0.2$ (i.e. a quadratic spline fit to the data) we can reduce the dispersion to $\sigma_z = 0.018$.

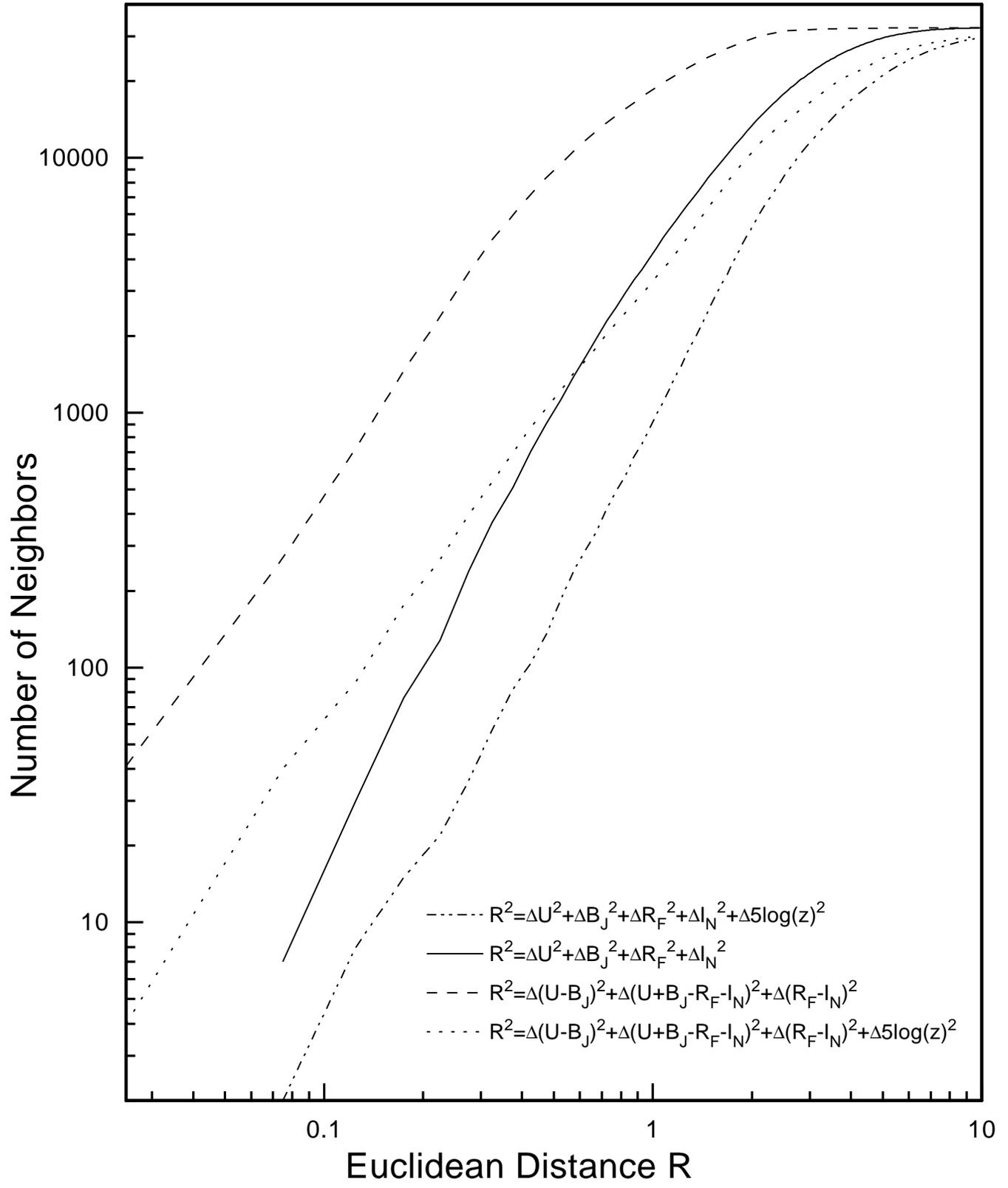

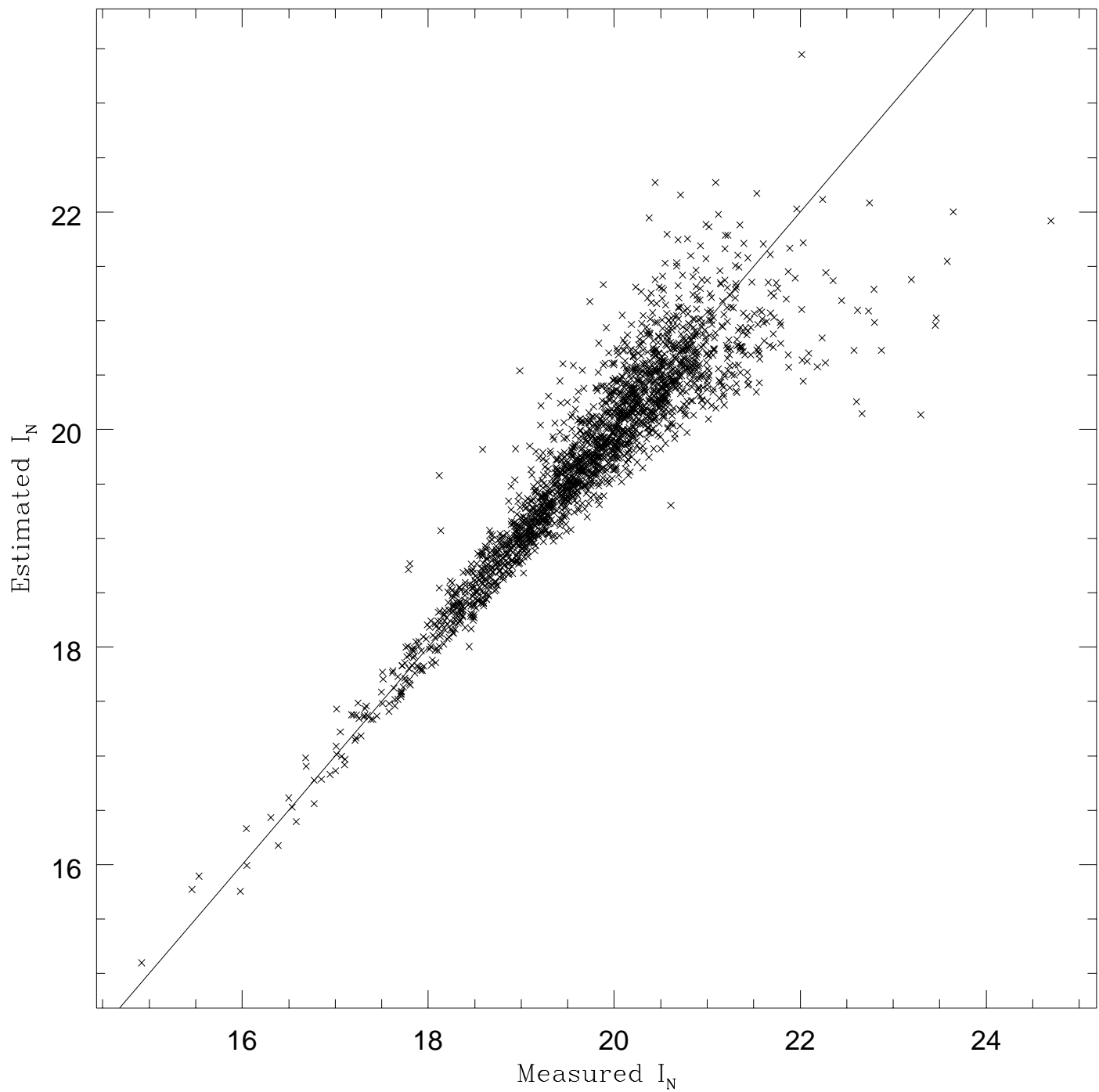

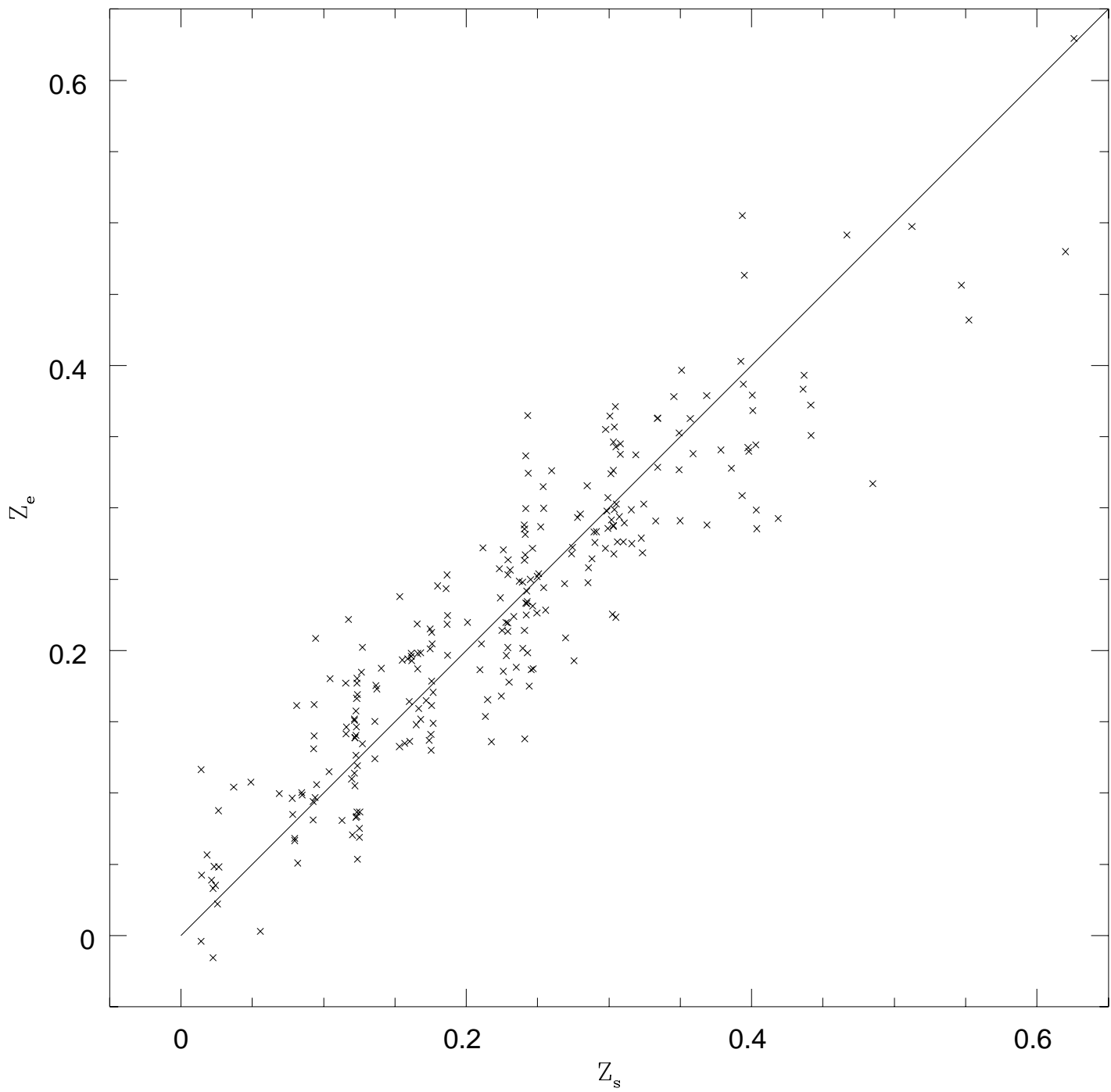

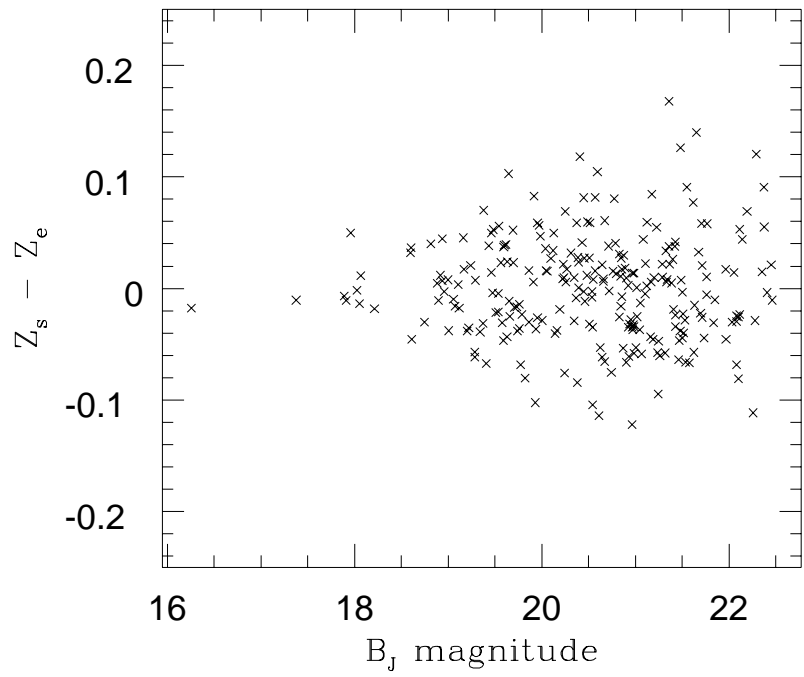
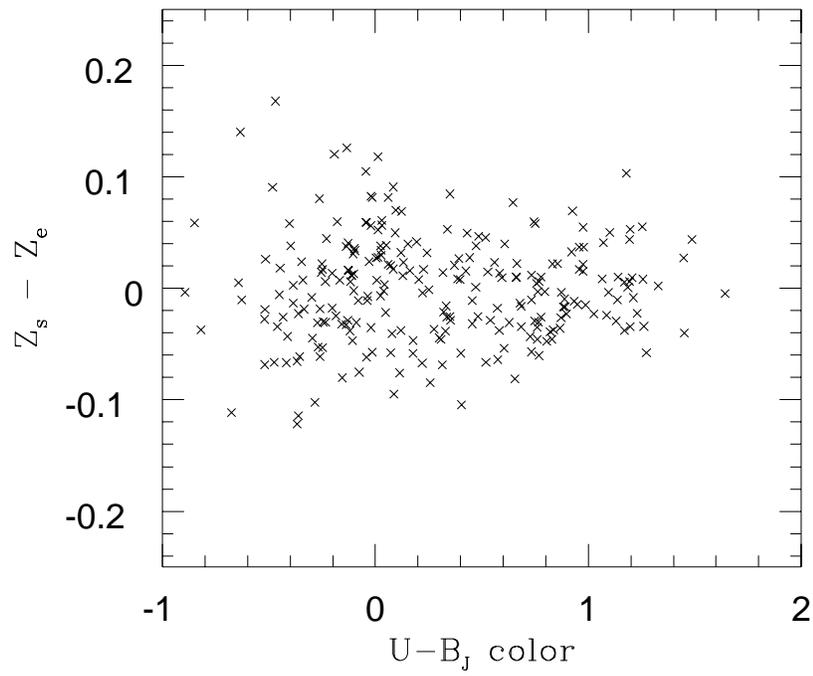

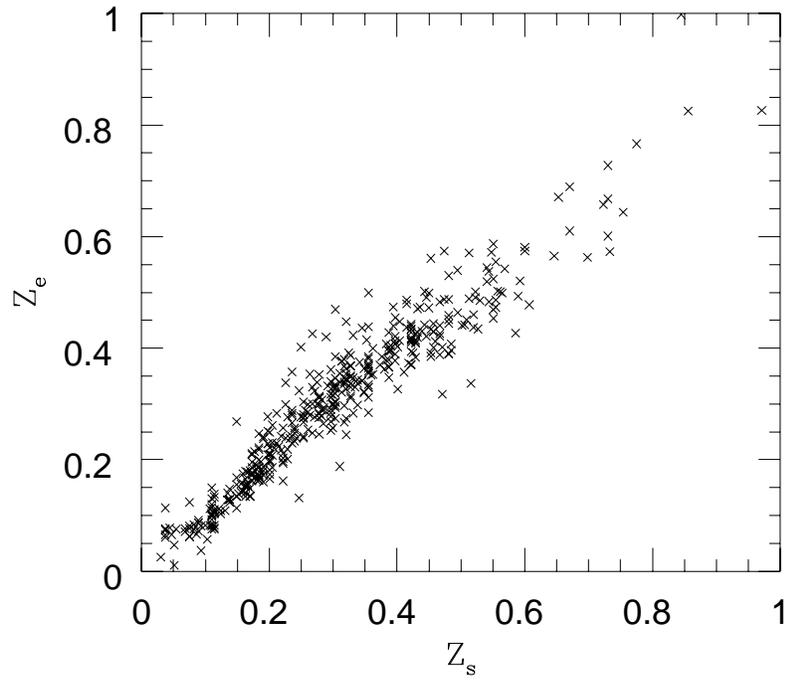

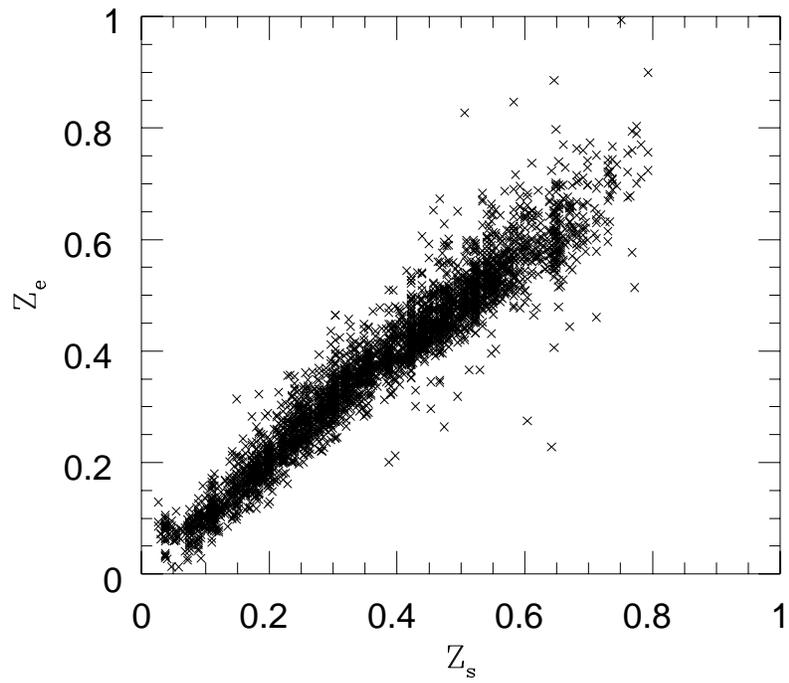